\documentclass[prd,preprint,aps,showpacs,tightenlines]{revtex4}

\usepackage{amsmath}

\renewcommand{\(}{\left(}
\renewcommand{\)}{\right)}
\renewcommand{\[}{\left[}
\renewcommand{\]}{\right]}
\renewcommand{\d}{\partial}
\newcommand{\f}[2]{\frac{#1}{#2}}
\newcommand{\df}[2]{\dfrac{#1}{#2}}
\newcommand{\pd}[3]{\left(\frac{\partial #1}{\partial #2}\right)_{#3}}

\newcommand{\xmu}{{x^{\mu}}}
\newcommand{\normal}{\hspace{-1mm}:\hspace{-1mm}}

\newcommand{\beq}{\begin{equation}}
\newcommand{\eeq}{\end{equation}}
\newcommand{\beqa}{\begin{eqnarray}}
\newcommand{\eeqa}{\end{eqnarray}}
\newcommand{\nol}{\nonumber \\ }

\begin{document}

\title{Radiation reaction on
charged particles in three-dimensional motion
in classical and quantum electrodynamics}

\author{Atsushi Higuchi$^1$ and Giles~D.~R.~Martin$^2$}

\affiliation{Department of Mathematics, University of York,
Heslington, York YO10 5DD, UK\\ email: $^1$ah28@york.ac.uk,
$^2$gdrm100@york.ac.uk}

\date{October 6, 2005}


\begin{abstract}
We extend our previous work (see arXiv:quant-ph/0501026),
which compared
the predictions of quantum electrodynamics concerning radiation
reaction with those of the Abraham-Lorentz-Dirac theory for
a charged particle in linear motion.
Specifically, we calculate the predictions for the change in
position of a charged
scalar particle, moving in three-dimensional space,
due to the effect of radiation reaction
in the one-photon-emission process in quantum electrodynamics.
The scalar particle is assumed to be accelerated for a finite
period of time by a three-dimensional
electromagnetic potential dependent only on one of the spacetime
coordinates.
We perform this calculation in the $\hbar\to 0$ limit and show that
the change in position agrees with that obtained
in classical electrodynamics with the Lorentz-Dirac force treated
as a perturbation. We also
show for a time-dependent but space-independent electromagnetic
potential that
the forward-scattering amplitude at order $e^2$
does not contribute to the position change
in the $\hbar \to 0$ limit after the mass renormalization is taken
into account.
\end{abstract}

\maketitle

\section{Introduction}

In classical electrodynamics, when a point charge $e$ with mass
$m$ is accelerated by a $4$-force $F^\mu_{\textrm{ext}}$, it
produces radiation and the equation of motion needs to be altered to
incorporate the radiation-reaction force. In the
Abraham-Lorentz-Dirac theory~\cite{Abraham,Lorentz,Dirac} we have
\begin{equation}
m\frac{d^2x^\mu}{d\tau^2} = F_{\rm ext}^\mu + F_{\rm LD}^\mu\,,
\label{LD}
\end{equation}
where $x^\mu$ are the spacetime coordinates of the charge at proper
time $\tau$ and where the
Lorentz-Dirac radiation-reaction force is given in the units $c=1$ by
\begin{equation}\label{LD4force}
F^\mu_{\rm LD} \equiv \frac{2\alpha_c}{3}\left[
\frac{d^3x^\mu}{d\tau^3} + \frac{dx^\mu}{d\tau}\left(\frac{d^2
x^\nu}{d\tau^2} \frac{d^2x_\nu}{d\tau^2}\right)\right]\,.
\end{equation}
Here we have defined $\alpha_c \equiv e^2/4\pi$, and our
metric convention is $g_{\mu\nu}={\rm diag}\,(+1,-1,-1,-1)$.
We assume in this paper
that the external force $F_{\rm ext}^\mu$ is a Lorentz force
resulting from a background electromagnetic field.
(See
Ref.~\cite{Teit} for an interesting derivation of this equation
and Ref.~\cite{Poisson} for a recent review.)
The Abraham-Lorentz-Dirac theory has a number of
problems if taken without modification, e.g. the existence
of unphysical run-away solutions in which the charge accelerates
under its own radiation. The suppression of this problem requires
the existence of acausal pre-acceleration.
(See, e.g., Refs.~\cite{Rohrlich,Jackson} for discussion of these
issues.)
However, the treatment
of the Lorentz-Dirac force as a perturbation, as used and justified
in Refs.~\cite{Landau,WE} for example,
is free of these difficulties.
Since classical electrodynamics
is now meaningful only as the $\hbar \to 0$ limit of quantum
electrodynamics (QED), which is the more fundamental theory,
it is natural
to ask how the perturbation theory of QED compares in the
classical limit to the Abraham-Lorentz-Dirac theory.
Indeed we
found recently that for a linearly accelerated charged particle
the Lorentz-Dirac force can be recovered from
the one-photon-emission amplitude in QED~\cite{Higuchi2,HM2,HM3}.
In this work we studied
the position of a charged-scalar wave packet in linear motion
and computed its change
due to the presence of radiation reaction in the
$\hbar \to 0$ limit.
The result was found to be in complete agreement
with that obtained by
using the Lorentz-Dirac force as a perturbation in classical
electrodynamics.

In this paper we generalize the work in Refs.~\cite{Higuchi2,HM2,HM3}
by extending the
calculations to a charged particle in three-dimensional motion
under an external electromagnetic
potential dependent on one of the spacetime coordinates.  We
also study the position change due to the forward-scattering amplitude
at order $e^2$. (Note that
Moniz and Sharp derived the Lorentz-Dirac force from QED
by considering a charge of
finite size and taking the zero-size limit~\cite{MS}.  Johnson and
Hu obtained the Lorentz-Dirac-like force for a
classical point charge
interacting with a massless scalar field by integrating out
the scalar field~\cite{beilok}.  See also Ref.~\cite{Tsyt} on the
relation between QED and the Lorentz-Dirac force.)
The structure of this paper is as follows. Sec.~\ref{clmodel}
outlines the model in the context of classical electrodynamics.
We then turn to the corresponding model for QED in
Sec.~\ref{qedmodel} and give the expressions for the position
shift of radiating particle relative to non-radiating particles in
terms of the one-photon-emission amplitude in the $\hbar \to 0$ limit.
In Sec.~\ref{emisamp} we calculate the emission amplitudes
for potentials dependent on one spacetime coordinate and then
proceed to calculate the quantum position shift in
Sec.~\ref{qposshift}. Then, Sec.~\ref{clposshift} gives an
expression for the classical position shift due to the addition of
a radiation-reaction force treated as a perturbation.  This quantity
is found to be identical with the $\hbar \to 0$ limit of the quantum
position shift.  We
summarize and discuss our results in Sec.~\ref{conclusion}. In
Appendix \ref{forscatappendix} we demonstrate that the forward-scattering
amplitude at order $e^2$ does not affect the
position of the particle in the $\hbar\to 0$ limit after the mass
renormalization is performed.

\section{Explanation of the model}\label{clmodel}

In this section we shall describe the model used for the
investigation in this article in the context of classical
electrodynamics. The model is the three-dimensional extension of
that used in Refs.~\cite{HM2,HM3} with the potential
dependent only on one of the space-time coordinates, $x^a$ say.  The particle is accelerated by an external force arising from
an electromagnetic potential $V^\mu$.  That is, the external force
in Eq.~(\ref{LD}) is given by
\beq
F^\mu_{\rm ext} = (\partial^\mu V^\nu - \partial^\nu V^\mu)
\frac{d x_\nu}{d\tau}\,.
\eeq
We wish to calculate the position shift, i.e. the change
in the position of the accelerated charge due to radiation
reaction, after a finite period of acceleration. Thus, we shall
assume that the potential $V^\mu$
behaves as follows: $V^\mu(x^a)=V_{(0)}^\mu$, where $V_{(0)}^\mu$ is
a constant vector, for $x^a<-X_1$ and $V^\mu(x^a)=0$ for $x^a>-X_2$
with $X_1$ and $X_2$ being positive constants ($X_1>X_2$). Hence,
the acceleration occurs only in the region $-X_1<x^a<-X_2$.
We also require that the coordinate $x^a$ of the particle increases
as a function of time. (This condition is automatically satisfied
if $x^a$ is the time coordinate.)
For the sake of simplicity we set
up the model so that the charged particle would pass the space
origin at time $t=0$ in the absence of radiation reaction.
(The spacetime origin is singled out as a
convenient point of reference because the quantum fields are
expanded in terms of the functions $e^{\pm ip\cdot x/\hbar}$,
which take the value one at the spacetime origin.  See
Ref.~\cite{HM3} for an analysis of the model in one dimension
with a reference point other than
the origin.)  The conditions on the external electromagnetic
potential imply that at $t=0$ the particle has already gone through the
region of acceleration.
The position shift is defined to be the change in
the space coordinates $x^i$ ($i=1,2,3$)
of this particle away from $x^i =0$
at $t=0$ due to radiation reaction.

It will be convenient for later use to write down the space components of
the Lorentz-Dirac force explicitly in terms of the velocity $v^i
\equiv \dot{x}^i$ and acceleration
$a^i \equiv \dot{v}^i$, where the dot represents
differentiation with respect to the
coordinate time $t$.
We write the equation of motion for
the Abraham-Lorentz-Dirac
theory (\ref{LD}) in a slightly different form as
\begin{equation}\label{eofm1}
\f{d\ }{dt}\left( m\,\frac{dx^i}{d\tau}\right)
= F_{\text{ext}}^i(x^a)\f{d\tau}{dt} + {\cal
F}_{{\rm LD}}^i\,.
\end{equation}
Then, defining ${\bf x}$, ${\bf v}$ and ${\bf a}$ to be
the vectors with $i$-th components $x^i$, $v^i$ and
$a^i$ respectively, and introducing the definition
$\gamma \equiv dt/d\tau =  (1-{\bf v}^2)^{-1/2}$
as usual, we obtain after a straightforward calculation
\begin{equation}\label{FcalLD}
{\cal F}_{LD}^i = \frac{2\alpha_c}{3} \left\{
  \f{d}{dt}\[\gamma^4 ({\bf a}\cdot {\bf v})v^i + \gamma^2 a^i\]
   -\gamma^6 ({\bf a}\cdot {\bf v})^2 v^i- \gamma^4 {\bf a}^2 v^i
\right\}\,.
\end{equation}

\section{Position-shift formula in QED}\label{qedmodel}

In this section we present the expression for the $\hbar\to 0$ limit of
the quantum position shift
in terms of the one-photon-emission amplitude.
The Lagrangian density for the quantum-field-theoretic model of a
charged scalar field $\varphi$ with mass $m$ and charge $e$
coupled to the electromagnetic field $A_\mu$ in the
Feynman gauge is given by
\begin{equation}\label{Lagrangian}
{\cal L} = [(D_\mu + ieA_\mu/\hbar)\varphi]^\dagger
(D^\mu +ieA^\mu/\hbar)\varphi -
(m/\hbar)^2\varphi^\dagger \varphi -
\frac{1}{4}F_{\mu\nu}F^{\mu\nu} - \frac{1}{2}(\partial_\mu A^\mu)^2\,,
\end{equation}
where $F_{\mu\nu} \equiv \partial_\mu A_\nu - \partial_\nu A_\mu$
and $D_\mu \equiv \partial_\mu + iV_\mu/\hbar$.
The function $V_\mu$ is the external potential which
accelerates the charged scalar particle as in the previous section.
For $e=0$ the charged scalar field is expanded as
\begin{equation}
\varphi(x) = \hbar \int \frac{d^3{\bf p}}{2p_0(2\pi\hbar)^3}
\left[ A({\bf p})\Phi_{\bf p}(x) + B^\dagger({\bf p}) \Phi_{\bf
p}^*(x)\right] \label{scalar_expansion}
\end{equation}
with $p_0 = \sqrt{{\bf p}^2 + m^2}$,
where the non-zero commutators are
\begin{equation}
\[ A({\bf p}), A^\dagger({\bf p'})\] = \[ B({\bf p}), B^\dagger({\bf p'})\]
= 2p_0 (2\pi\hbar)^3 \delta^3({\bf p}-{\bf p'})\,.
\end{equation}
The mode functions $\Phi_{{\bf p}}(x)$ satisfy the Klein-Gordon
equation in the presence of the external field $V_\mu$:
\beq
(D^\mu D_\mu + m^2)\Phi_{{\bf p}}(x) = 0\,. \label{KGeq}
\eeq
If the potential is purely $t$-dependent, then we require that
$\Phi_{{\bf p}}(x)$ take the form $e^{-ip\cdot x/\hbar}$, where
$p\cdot x \equiv p^0 t- {\bf p}\cdot {\bf x}$,
near the hypersurface $t=0$.  (This
is possible because the potential $V_\mu$ vanishes
there by assumption.)  If the
potential is space-dependent, then we require
that $\Phi_{{\bf p}}(x)$ take the form $e^{-ip\cdot x/\hbar}$
in a region including the spacetime origin~\footnote{ This
choice may not be possible for all ${\bf p}$ if $V^\mu_{(0)}\neq
0$ because, then, there may be modes which are completely
reflected; the expansion (\ref{scalar_expansion}) will have to be
modified for these modes. This modification will not be relevant
to our calculations since the modes for which the WKB
approximation is valid --- the only modes we are interested in --- are not
affected by this modification.}. Similarly, the electromagnetic
field is expanded as
\begin{equation}
A_\mu(x) = \int\frac{d^3{\bf k}}{2k(2\pi)^3} \left[ a_\mu({\bf k})
e^{-ik\cdot x} + a^\dagger_\mu({\bf k})e^{ik\cdot x} \right]
\end{equation}
with $k = \|{\bf k}\|$ and
$k\cdot x \equiv kt - {\bf k}\cdot {\bf x}$,
where the annihilation and creation operators,
$a_\mu({\bf k})$ and $a^\dagger_\mu({\bf k})$,
for the photons with momenta $\hbar {\bf k}$ satisfy
\begin{equation}
\left[ a_\mu({\bf k}),a^\dagger_\nu({\bf k}')\right] =
-g_{\mu\nu}(2\pi)^32\hbar k\delta^3({\bf k}-{\bf k}')\,.
\end{equation}
Notice that the scalar field $\varphi$
is expanded in terms of the momentum ${\bf p}$ whereas
the electromagnetic field $A_\mu$ is expanded
in terms of the wave number ${\bf k}$.
We adopt this convention because the vectors
${\bf p}$ and ${\bf k}$ are regarded as classical
rather than ${\bf p}/\hbar$, the wave number
of the scalar particle, and $\hbar{\bf k}$, the momentum of the
electromagnetic field.

Let the initial wave-packet state be given by
\begin{equation}\label{istate}
| i\rangle = \int \frac{d^3{\bf p}}{\sqrt{2p_0}(2\pi\hbar)^3}
f({\bf p})A^\dagger({\bf p})|0\rangle\,,
\end{equation}
where the function $f({\bf p})$ is sharply peaked about a given
momentum with width of order $\hbar$.
If the potential depends on a space coordinate $x^a$, then we
require that the $a$-th component of this momentum be positive
so that the wave packet is near the spacetime origin
{\em after} it has undergone acceleration.  The
normalization of the operators $A^\dagger({\bf p})$ is such that
the condition $\langle i\,|\,i\rangle = 1$ leads to
\begin{equation}\label{inorm}
\int\frac{d^3{\bf p}}{(2\pi\hbar)^3}|f({\bf p})|^2 = 1\,.
\end{equation}
This shows that the function $f({\bf p})$ can heuristically
be regarded as the
one-particle wave function in the momentum representation.
The position expectation value is given by
\begin{equation}
\langle x^i\rangle = \int d^3{\bf x}\, x^i\langle \rho({\bf x},t)
\rangle \, , \label{zedd}
\end{equation}
where the density operator for the scalar field is
\begin{equation}
\rho(x) \equiv\frac{i}{\hbar}\normal \varphi^\dagger\partial_t \varphi -
\partial_t \varphi^\dagger\cdot \varphi\normal\,. \label{charge}
\end{equation}
In Ref.~\cite{HM3} it was shown that for the non-radiating particle the
position expectation value at $t=0$ is given to lowest order by
the expression
\begin{equation}\label{nonradshift}
\langle x^i \rangle_0 = \frac{i\hbar}{2} \int\frac{d^3{\bf
p}}{(2\pi\hbar)^3} f^*({\bf p})
\stackrel{\leftrightarrow}{\partial}_{p^i} f({\bf p}) \, ,
\end{equation}
where $\stackrel{\leftrightarrow}{\partial}_{p^i} =
\stackrel{\rightarrow}{\partial}_{p^i} - \stackrel{\leftarrow}
{\partial}_{p^i}$. This expression can heuristically be regarded as the
expectation value of the position operator $i\hbar\partial_{p^i}$ in
the momentum representation.  It also agrees
with the expectation value
of the Newton-Wigner operator~\cite{Newton-Wigner}.
We let the function $f({\bf p})$ satisfy $\langle x^i \rangle_0=0$
so that we measure the position shift for the particle relative to
the origin as we required in the previous section.

For the radiating particle,
the evolution to order $e^2$ in perturbation theory in interaction
picture takes the following form:
\begin{eqnarray}
A^\dagger({\bf p})|0\rangle  \to  [1 + i{\cal F}({\bf
p})]A^\dagger({\bf p})|0\rangle + \frac{i}{\hbar} \int
\frac{d^3{\bf k}}{(2\pi)^3 2k} {\cal A}^\mu({\bf p},{\bf k})
a_\mu^\dagger({\bf k})A^\dagger({\bf P})|0\rangle\,, \label{trans}
\end{eqnarray}
where ${\cal F}({\bf p})$ is the forward-scattering amplitude
and where ${\cal A}^\mu({\bf p},{\bf k})$
is the amplitude
for the one-photon emission.  If the potential is purely $t$-dependent,
then the momentum is conserved and one has
${\bf P} = {\bf p}-\hbar{\bf k}$.
If the potential is $x^3$-dependent, say, then the momentum
${\bf P}$
is determined by energy conservation
$\sqrt{{\bf p}^2 + m^2} = \sqrt{{\bf P}^2 + m^2} +
\hbar k$ and transverse-momentum conservation
$p^i = P^i + \hbar k^i$ ($i=1,2$).
Thus, the final state
for the initial wave-packet state $|i\rangle$ given
by Eq.~(\ref{istate}) is
\begin{eqnarray}\label{final1}
| f\rangle & = & \int\frac{d^3{\bf p}}{\sqrt{2p_0}(2\pi\hbar)^3}
\left[ 1 + i{\cal F}({\bf p})\right]
f({\bf p}) A^\dagger({\bf p})|0\rangle
\nonumber
\\
&& + \frac{i}{\hbar}\int \frac{d^3{\bf k}}{2k(2\pi)^3}
\int\frac{d^3{\bf p}}{\sqrt{2p_0}(2\pi\hbar)^3} {\cal A}^\mu({\bf
p},{\bf k})f({\bf p}) a_\mu^\dagger({\bf k})A^\dagger({\bf
P})|0\rangle\,.
\end{eqnarray}
The position expectation value for this state in the
$\hbar \to 0$ limit
can be calculated by using Eq.~(\ref{zedd}) and
was given in Ref.~\cite{HM3}.  It consists of
three terms identified as the non-radiating
position expectation value, $\langle x^i\rangle_0$, which we have set to
zero for simplicity, the forward-scattering contribution
(which is not due to the radiation) and a third term
identified as the position shift due to radiation reaction,
which we denote by $\delta x^i_Q$.
Thus, the position expectation value at $t=0$ is
\beq \label{forwardadded}
\langle x^i\rangle_{t=0}
= - \hbar\frac{\partial\ }{\partial p^i}{\rm Re}\,{\cal F}({\bf p})
+ \delta x^i_Q
\eeq
with
\begin{equation}\label{shiftA}
\delta x^i_Q = - \frac{i}{2} \int \frac{d^3{\bf
k}}{2k(2\pi)^3} {\cal A}^{\mu *}({\bf p},{\bf k})
\stackrel{\leftrightarrow}{\partial}_{p^i} {\cal A}_\mu({\bf
p},{\bf k})
\end{equation}
in the $\hbar\to 0$ limit, where the wave packet becomes narrowly
peaked about the momentum ${\bf p}$.
({}From now on, the momentum ${\bf p}$ will always
refer to the peak value for the wave-packet state.)
Thus, the position shift $\delta x_Q^i$ can be found from
the emission amplitude ${\cal A}_\mu({\bf p},{\bf k})$
to be calculated for a given external electromagnetic potential that accelerates the particle.
Although this expression was considered only
for linear acceleration in Ref.~\cite{HM3}, it is equally valid for
motion in three dimensions: the details of the potential
and the path of the particle are unnecessary for its derivation.

It will
be shown in Appendix A that the forward-scattering contribution in
Eq.~(\ref{forwardadded}) vanishes in the $\hbar \to 0$ limit
after the mass renormalization is taken into account if the potential
depends only on time, and we expect that this is also the
case for potentials dependent on one space coordinate.
For this reason
we concentrate on the evaluation of the radiation-reaction
contribution $\delta x^i_Q$ in the rest of this paper.

\section{Emission Amplitude}\label{emisamp}

The photon-emission part of the evolution of the state in
Eq.~(\ref{trans}) is given
in terms of the emission amplitude by
\begin{equation}
A^\dagger({\bf p})|0\rangle \to \cdots + \frac{i}{\hbar}
\int\frac{d^3{\bf k}}{2k(2\pi)^3} {\cal A}^{\mu}({\bf p},{\bf
k})a_\mu^\dagger({\bf k}) A^\dagger({\bf P})|0\rangle \, .
\label{evolA}
\end{equation}
The evolution of this state to first order in time-dependent
perturbation theory is
\begin{equation}\label{evolL}
A^\dagger({\bf p})|0\rangle \to \cdots -\frac{i}{\hbar}\int d^4 x
{\cal H}_I(x)A^\dagger({\bf p})|0\rangle\,,
\end{equation}
where ${\cal H}_I(x)$ is the interaction Hamiltonian density.
Comparing these two evolution expressions, we can write the emission
amplitude in terms of the interaction Hamiltonian density as
\begin{equation}
{\cal A}_\mu ({\bf p},{\bf k}) =
\frac{1}{\hbar}\int \frac{d^3{\bf p'}}{2p_0'(2\pi\hbar)^3} \int
d^4x\langle 0|a_\mu({\bf k})A({\bf p'}){\cal H}_I(x)A^\dagger({\bf
p})|0\rangle\,.
\end{equation}
The Hamiltonian density
can be obtained from the Lagrangian density
(\ref{Lagrangian}) by the standard procedure.  Thus, we find
the following interaction Hamiltonian density:
\begin{equation}\label{Hamiltonian}
{\cal H}_I(x) = \frac{ie}{\hbar}A_\mu \normal\left[ \varphi^\dagger
D^\mu
\varphi - (D^\mu \varphi)^\dagger \varphi\right]\normal
 + \frac{e^2}{\hbar^2}\sum_{i=1}^3 A_iA_i
\normal\varphi^\dagger\varphi\normal - \frac{\delta m^2}{\hbar^2}
\normal\varphi^\dagger\varphi\normal\,,
\end{equation}
where $D_\mu\equiv \partial_\mu + iV_\mu/\hbar$ as before.
We have normal-ordered the scalar-field operators
to drop the vacuum polarization
diagram automatically. (Note that the second term is different
from what might be na\"{\i}vely expected,
$-(e^2/\hbar^2)A_\mu A^\mu\normal\varphi^\dagger \varphi\normal\,\,$.
This difference
is due to the presence of interaction terms involving
$\dot{\varphi}$ or $\dot{\varphi}^\dagger$
in the Lagrangian density.) The last term in Eq.~(\ref{Hamiltonian})
is the mass counterterm.  The counterterm for the wave-function
renormalization will not be necessary in our calculations.
A photon is
emitted due to the first term in the Hamiltonian density
(\ref{Hamiltonian}).  Thus, we have
\begin{eqnarray}
{\cal A}_\mu ({\bf p},{\bf k}) & = & \frac{ie}{\hbar^2}\int
\frac{d^3{\bf p'}}{2p_0'(2\pi\hbar)^3} \int d^4x \langle
0|a_\mu({\bf k})A({\bf p'})\left\{ A_\nu \normal\left[ \varphi^\dagger
D^\nu \varphi - (D^\nu\varphi)^\dagger \varphi\right]\normal \right\}
A^\dagger({\bf p})|0\rangle\,. \nonumber \\
\end{eqnarray}
By using the expansion of the fields $A_\mu$ and $\varphi$, and the
commutation relations
for the annihilation and creation operators, one readily
finds
\begin{eqnarray}\label{emisampPhi}
 {\cal A}_\mu ({\bf p},{\bf k}) &=& -ie\hbar \int \frac{d^3{\bf
p'}}{2p_0'(2\pi\hbar)^3} \int d^4x\,  e^{i k\cdot x}
 \left\{ \Phi^*_{{\bf p}'}(x) D_\mu \Phi_{\bf p}(x) - \left[ D_\mu
\Phi_{{\bf p}'}(x)\right]^\dagger \Phi_{\bf p}(x) \right\}\,.
\end{eqnarray}

Next we shall find a simple expression for
the $\hbar\to 0$ limit of the
emission amplitude given by Eq.~(\ref{emisampPhi})
for an electromagnetic potential which
depends only on one spacetime coordinate.  We present the details of the
calculation only for a $t$-dependent potential.  The final result for
an $x$- $y$- or $z$-dependent potential is identical, and the techniques
involved are mostly the same.
Thus, we consider a time-dependent potential $V^i(t)$ ($i=1,2,3$)
with  $V^0(t) = 0$.  [If $V^0(t) \neq 0$, one can gauge away this
component.]
The system is translationally invariant in the
spatial directions and, hence, we can let
\begin{equation}
\Phi_{{\bf p}}({\bf x},t) = \phi_{\bf p}(t)\exp\(i{\bf p}\cdot{\bf
x}/\hbar\)\,.
\label{Phiphi}
\end{equation}
The amplitude in a spatial direction for the $t$-dependent
potential is then
\beqa
{\cal A}^i({\bf p},{\bf k})
 &=& -e\int\f{d^3{\bf p}'}{2p_0'(2\pi\hbar)^3}\int d^4x \,
 \phi^*_{\bf p'}(t)\phi_{\bf p}(t) \[p^i+p^{\prime i}
-2V^i(t)\]e^{i\[({\bf p}-{\bf p'})
\cdot {\bf x}/\hbar\]}e^{i(k t - {\bf k}\cdot{\bf x})}
\nol
&=& -e\int dt \, e^{ikt} \phi^*_{\bf P}(t)\phi_{\bf p}(t)
\f{p^i-V^i(t)}{p_0}\label{inter1}
\eeqa
with ${\bf P} = {\bf p}-\hbar {\bf k}$, where we have let
$P^i + p^i -2V^i(t) = 2[p^i-V(t)]$ as
the difference $p^i - P^i$ is of order
$\hbar$~\footnote{In Appendix A the semiclassical approximation for
the emission probability (\ref{imaginary}) is justified.  This shows
the validity of the physically reasonable
assumption that a typical photon emitted
has energy of order $\hbar$.}.
It would be wrong to equate $\phi_{\bf P}(t)$ with $\phi_{\bf p}(t)$
because
these functions oscillate with periods of order $\hbar^{-1}$.
For the time component we have $D_t=\d_t$ and simply obtain
\beq
{\cal A}^0({\bf p},{\bf k}) =
-\f{ie\hbar}{2p_0}\int dt\,\[\phi^*_{\bf P}\d_t\phi_{\bf p} -
 \(\d_t\phi^*_{\bf P}\)\phi_{\bf p}\] e^{ikt}\,.\label{inter2}
\eeq

To proceed further we need to approximate the
function $\phi_{\bf p}(t)$
in a way suitable for taking
the $\hbar \to 0$ limit.  To this end we use
the semi-classical WKB approximation.
By substituting Eq.~(\ref{Phiphi}) in the Klein-Gordon equation
(\ref{KGeq}) we find
\beq\label{phiequation}
\left\{\hbar^2 \d_t^2 + \[{\bf p} - {\bf V}(t)\]^2 +m^2\right\}
\phi_{\bf p}(t)=0\,,
\eeq
where ${\bf V}(t)$ is the vector with its $i$-th component given by
$V^i(t)$.
The standard WKB approximation gives the following positive-frequency
solution:
\beqa\label{wkbt}
\phi_{\bf p}(t) & = &
 \sqrt{\f{p_0}{\sigma_{\bf p}(t)}}\exp\[-\f{i}{\hbar}\int^t_0
\sigma_{\bf p}(\zeta)d\zeta
 \] \,,
\eeqa
where
\beq
\sigma_{\bf p}(t) \equiv
\sqrt{\[{\bf p} - {\bf V}(t)\]^2+m^2}\,.
\eeq
We note that the local momentum and energy of the point particle
corresponding to the wave packet considered here are
\beqa
m\frac{d{\bf x}}{d\tau} & = & {\bf p} - {\bf V}(t)\,,\label{momentum1}\\
m\frac{dt}{d\tau} & = & \sigma_{\bf p}(t)\,.\label{energy1}
\eeqa
Now, the product of two wave functions
in the emission amplitude (\ref{inter1}) can be written
\beq
\phi^*_{\bf P}(t)\phi_{\bf p}(t) =
\f{p_0}{\sigma_{\bf p}(t)}
\exp\left\{-\f{i}{\hbar}\int^t_0
\[\sigma_{\bf p}(\zeta)-\sigma_{\bf P}(\zeta)\]d\zeta\right\}\,,
\label{phiproduct}
\eeq
where we have replaced
$P_0$ and $\sigma_{\bf P}(t)$ in the pre-factor by
$p_0$ and $\sigma_{\bf p}(t)$, respectively, because
we are interested only in the $\hbar \to 0$ limit.
The integrand in the exponent can be evaluated to lowest order in
$\hbar$ by using
Eqs.~(\ref{momentum1}) and (\ref{energy1}) as
\beqa
\sigma_{\bf p}-\sigma_{\bf P} &=& \f{\d \sigma_{\bf p}}{\d
p^i}(P^i-p^i) \nol
 &=& \f{dx^i}{dt}\hbar k^i\,, \label{wkbvel}
\eeqa
where the repeated indices $i$ are summed over. By substituting this
approximation in
Eq.~(\ref{phiproduct}) we find
\beqa
\phi^*_{\bf P}(t)\phi_{\bf p}(t) &=&
\f{p_0}{\sigma_{\bf p}}\exp\(-i \int^t_0 dt\, \f{dx^i}{dt}
k^i\) \nol
 &=& \f{p_0}{\sigma_{\bf p}}\exp\(-i{\bf k}\cdot {\bf x}\)\,,
\label{phipronice}
\eeqa
where we have used the fact that the particle passes through the
spacetime origin.
By substituting
 this formula in Eq.~(\ref{inter1}) and noting
Eqs.~(\ref{momentum1}) and (\ref{energy1}) we obtain
\beqa
{\cal A}^i({\bf p},{\bf k}) &=&
- e\int dt\, e^{ik\cdot x}\f{dx^i}{dt}\nol
 &=& - e \int d\xi\, \f{dx^i}{d\xi} e^{ik\xi}\,, \label{Aiai}
\eeqa
where we have defined
$\xi\equiv t-{\bf n}\cdot {\bf x}$ with ${\bf n} \equiv {\bf k}/k$\,.
We emphasize that ${\bf x}$ and $\xi$ here
are functions of $t$ evaluated
on the world line of the corresponding classical particle passing
through the spacetime origin.

Let us now consider the time component ${\cal A}^0({\bf p},{\bf k})$
of the emission amplitude given by Eq.~(\ref{inter2}).
Note that from the WKB expression (\ref{wkbt}) for $\phi_{\bf p}(t)$
we have, to lowest order in $\hbar$,
\beq
\d_t\phi_{\bf p}(t) = -\f{i}{\hbar}\sigma_{\bf p}(t)\phi_{\bf p}(t)\,.
\eeq
By substituting this formula in Eq.~(\ref{inter2}) we obtain
\beqa
{\cal A}^0({\bf p},{\bf k})
&=& -\f{e}{2p_0}\int dt\, \left[\sigma_{\bf p}(t) +
\sigma_{\bf P}(t)\right]
\phi^*_{\bf P}(t)\phi_{\bf p}(t)e^{ikt} \nol
 &=&-e\int dt\,e^{-i{\bf k}\cdot {\bf x}}e^{ikt} \nol
 &=& -e \int d\xi\, \f{dt}{d\xi} e^{ik\xi}\,,
\eeqa
where we have let $\sigma_{\bf P}(t)=\sigma_{\bf p}(t)$ and used
Eq.~(\ref{phipronice}).
By combining this formula and Eq.~(\ref{Aiai}) we obtain the following
concise expression for the $\hbar \to 0$ limit of the emission amplitude:
\beq\label{amplitude}
{\cal A}^\mu({\bf p},{\bf k})
= -e \int d\xi\, \f{dx^\mu}{d\xi} e^{ik\xi}\,.
\eeq
One can proceed in a similar manner for the case
where the potential depends on any
one of the spatial coordinates. One obtains Eq.~(\ref{amplitude})
in this case as well. The emission amplitude (\ref{amplitude})
is identical to that for a
classical particle corresponding to the wave packet
considered here~\cite{HM2}.

\section{Quantum position shift}\label{qposshift}

The expression (\ref{amplitude})
for the emission amplitude is
ill-defined, as noted in Ref.~\cite{HM3} for the one-dimensional case,
because the integrand does not tend to zero as
$\xi \to \pm \infty$. For this reason
we introduce a smooth cut-off function $\chi(\xi)$ which takes the
value one
while the acceleration is non-zero and has the property
$\lim_{\xi\to\pm\infty}\chi(\xi) = 0$, viz.
\begin{equation}
{\cal A}^\mu({\bf p},{\bf k})  =  -e \int_{-\infty}^{+\infty}
d\xi\, \frac{dx^\mu}{d\xi}\chi(\xi)\,e^{ik\xi}\,.\label{cut-off}
\end{equation}
One can show exactly as in Ref.~\cite{HM3} that
the substitution of this
expression into the quantum position shift given by
Eq.~(\ref{shiftA}) leads to an expression independent of the cut-off
function as follows:
\begin{equation}\label{quantshifteqn}
  \delta x^i_{\rm Q} = -\frac{\alpha_c}{4\pi}\int d\Omega \int dt\,
\frac{d^2 x_{\mu}}{d\xi^2}
\frac{d\ }{dt}\left(\frac{\partial x^\mu}{\partial p^i}
\right)_{\xi}\,,
\end{equation}
where $d\Omega$ is the solid angle in the
wave-number space of the photon emitted.
The variable appearing as the subscript for
a partial derivative --- the variable $\xi$ in this equation --- is held
fixed. [Note that
$dx^\mu/dt = \left(\partial x^\mu/\partial t\right)_{p^i}$.]

To evaluate $\delta x_Q^i$ in Eq.~(\ref{quantshifteqn}) we first
need to put the integrand in the form amenable to the solid-angle
integration by rewriting it in terms of $t$ instead of $\xi$.
One can readily write $d^2 x^\mu/d\xi^2$ in terms of
$t$-derivatives by using $d/d\xi = (1-n^i\dot{x}^i)^{-1}d/dt$
as follows:
\beq\label{d2xdxi2}
\f{d^2\xmu}{d\xi^2} = \dot{\xi}^{-3}
\[\(1-n^i\dot{x}^i\)\ddot{x}^\mu +
n^i\ddot{x}^i\dot{x}^\mu\]\,,
\eeq
where $\dot{\xi} = 1 - n^i \dot{x}^i$.
Here and in the rest of this section,
Latin indices take the spatial values 1 to 3, and are summed over
when repeated.
The time and space components of Eq.~(\ref{d2xdxi2})
can separately be given as
\beqa
\f{d^2t}{d\xi^2} &=& \dot{\xi}^{-3}\,n^i\ddot{x}^i\,,
\label{secondder1}\\
\f{d^2x^j}{d\xi^2} &=&\dot{\xi}^{-3}
\[\(1-n^i\dot{x}^i\)\ddot{x}^j +
n^i\ddot{x}^i\dot{x}^j\]\,.
\label{secondder2}
\eeqa
Next we express $(\partial x^\mu/\partial p^i)_\xi$ in
Eq.~(\ref{quantshifteqn}) in the form
involving $t$ rather than $\xi$
as follows.  Note first
\begin{equation}
dx^\mu =
\frac{dx^\mu}{dt}dt + \left(\frac{\partial x^\mu}
{\partial p^i}\right)_t dp^i\,.
\end{equation}
[The zeroth component of this equation is trivial because
$(\partial t/\partial p^i)_t = 0$.]  By substituting $dt=d\xi + n^kdx^k$
in this equation
with $\mu=j$ and solving for $dx^i$, we obtain
\begin{equation}
dx^i = \frac{v^i}{1-n^jv^j}d\xi +
\frac{[\delta^{ik}(1-n^lv^l)+n^kv^i]}{1-n^lv^l}
\left(\frac{\partial x^k}{\partial p^j}\right)_t dp^j\,,
\label{partial1}
\end{equation}
where $v^i \equiv \dot{x}^i$.
Hence
\begin{equation}
\left(\frac{\partial x^i}{\partial p^j}\right)_\xi
= \frac{[\delta^{ik}(1-n^lv^l)+n^kv^i]}{1-n^lv^l}
\left(\frac{\partial x^k}{\partial p^j}\right)_t\,.
\label{partial2}
\end{equation}
With $\xi$ fixed we have $dt - n^i dx^i = d\xi = 0$. Thus,
\begin{eqnarray}
\left(\frac{\partial t}{\partial p^j}\right)_\xi
& = & n^i \left(\frac{\partial x^i}{\partial p^j}\right)_\xi\nol
& = &
\frac{n^i}{1-n^l v^l}\left(\frac{\partial x^i}{\partial p^j}
\right)_t\,. \label{partial3}
\end{eqnarray}

By substituting Eqs.~(\ref{secondder1}), (\ref{secondder2}),
(\ref{partial2}) and (\ref{partial3}) in Eq.~(\ref{quantshifteqn}) we
find the following expression after a tedious but straightforward
calculation:
\beqa
\delta x^i_Q
 &=& -\frac{\alpha_c}{4\pi}\int dt\left\{
\[ I_2^{kj} \gamma^{-2}a^j
- I_0 a^k - I_1^j a^j v^k - I_1^k ({\bf a}\cdot
{\bf v})
\] \frac{d\ }{dt}\left(\frac{\partial x^k}{\partial p^i}\right)_t
\right. \nonumber \\
&&
\ \ \ \ \ \ \ \ \ \ \ \ \ \ \ \left. + \[
I_3^{kjl}\gamma^{-2}a^ja^l
- 2I_2^{kj}a^j({\bf a}\cdot {\bf v})
 - I_1^k {\bf a}^2  \] \left(\frac{\partial x^k}{\partial p^i}
\right)_t\right\}\,,  \label{almostthere}
\eeqa
where
\beqa
I_0 &\equiv & \displaystyle\int d\Omega \df{1}{\dot{\xi}^2}
= 4\pi\gamma^2\,, \label{integral1} \\
I_1^i &\equiv & \displaystyle\int d\Omega \df{n^i}{\dot{\xi}^3}
= 4\pi\gamma^4 v^i\,, \\
I_2^{ij} &\equiv & \displaystyle\int d\Omega
\df{n^i n^j}{\dot{\xi}^4}
=  \f{16}{3}\pi\gamma^6 v^i v^j   +
\f{4}{3}\pi\gamma^4 \delta^{ij}\,, \\
I_3^{ijk} &\equiv & \displaystyle\int d\Omega
\df{n^in^jn^k}{\dot{\xi}^5} =  8\pi\gamma^8 v^iv^jv^k +\f{4}{3}\pi\gamma^6\(
v^i\delta^{jk}
 +v^j\delta^{ik}+v^k\delta^{ij} \)\,. \label{integral4}
\eeqa
Evaluation of these
solid-angle integrals
is facilitated by noting that the last three integrals
are proportional to partial derivatives of $I_0$ with respect to $v^i$.
Substitution of Eqs.~(\ref{integral1})--(\ref{integral4}) in
Eq.~(\ref{almostthere}) yields
\beq
\delta x^i_Q = \f{2\alpha_c}{3}\int dt \left\{
\[\gamma^4 ({\bf a}\cdot {\bf v}) v^k + \gamma^2 a^k \] \f{d\ }{dt} \(
\f{\partial x^k}{\partial p^i}\)_t +
\[\gamma^6 ({\bf a}\cdot {\bf v})^2 v^k +\gamma^4{\bf a}^2 v^k  \] \(
\f{\partial x^k}{\partial p^i}\)_t \right\}\,.  \label{xiQOK}
\eeq
By comparing this equation with the expression (\ref{FcalLD}) for the
Lorentz-Dirac force,
we note that the quantum position shift can be written as
\beq \label{possposs}
\delta x^i_Q = - \int^0_{-\infty}
dt\, {\bf {\cal F}}^{j}_{\rm LD} \pd{x^j}{p^i}{t}\,.
\eeq
We have used the
fact that ${\bf a}(t) \neq 0$ only for $t < 0$ and
for a finite interval of time to integrate the first term in
Eq.~(\ref{xiQOK}) by parts.
In the next section we shall demonstrate that this expression for the
position shift agrees with that in the Abraham-Lorentz-Dirac theory.
We recall here
that $p^i$ is the momentum of the particle at $t=0$, which
is after it has gone through the acceleration.

\section{Classical position shift}\label{clposshift}

In this section we discuss the classical perturbative solution to the
Lorentz-Dirac equation (\ref{eofm1}) in general terms in order to
show that the classical and quantum position shifts coincide.
We do not need the assumption that the external electromagnetic
potential depends only on one coordinate in this section;
this assumption was necessary
only for the calculation of the quantum position shift $\delta x_Q^i$.

Recall that the classical motion of a charge $e$ with mass $m$ in an
electromagnetic potential $V^\mu = (V^0, {\bf V})$,
which depends on $t$ and ${\bf x}$, is described by the
following Hamiltonian:
\begin{equation}
H = \sqrt{\( {\bf P} - {\bf V}\)^2 +m^2} + e V^0\,,
\end{equation}
where ${\bf P}$ is the momentum conjugate to ${\bf x}$.
One can readily show that the Lorentz-Dirac equation
(\ref{eofm1}) is equivalent to the
following set of equations:
\beqa
\dot{x}^i & = & \f{\partial H}{\partial P^i}\,,\\
\dot{P}^i & = & -\f{\partial H}{\partial x^i} + {\cal F}_{\rm LD}^i\,.
\eeqa
Now, let $({\bf x},{\bf P}) = ({\bf x}_0(t),{\bf P}_0(t))$ be the
solution
to these coupled equations with ${\cal F}_{\rm LD}^i=0$ satisfying
${\bf x}_0(0)=0$ and ${\bf P}_0(0) = {\bf p}$.  This solution
gives the classical trajectory of the particle passing through
the spacetime origin with final momentum ${\bf p}$ in the absence
of radiation reaction.
If $({\bf x},{\bf P}) = ({\bf x}_0(t) + \delta {\bf x}(t),
{\bf P}_0(t) + \delta {\bf P}(t))$ is the retarded solution
to these equations with ${\cal F}_{\rm LD}^i \neq 0$ to first order in
${\cal F}_{\rm LD}^i$, then $\delta{\bf x}$ and $\delta {\bf P}$ will
have the property that
$(\delta {\bf x},\delta {\bf P}) \to (0,0)$ as $t\to -\infty$ and satisfy
\beqa
\f{d\ }{dt}\delta x^i & = & \f{\partial^2 H}{\partial x^j\partial P^i}
\delta x^j + \f{\partial^2 H}{\partial P^j\partial P^i}\delta P^j\,,
\label{LDone}\\
\f{d\ }{dt}\delta P^i & = & -\f{\partial^2 H}{\partial x^j\partial x^i}
\delta x^j
- \f{\partial^2 H}{\partial x^i P^j} \delta P^j + {\cal F}_{\rm LD}^i\,,
\label{LDtwo}
\eeqa
where the partial derivatives of $H$ are evaluated at
$({\bf x},{\bf P}) = ({\bf x}_0(t),{\bf P}_0(t))$ and, hence, are
functions of $t$ alone.  The quantity
$\delta x_C^i \equiv \delta x^i(0)$
is identified as the position shift in the classical
Abraham-Lorentz-Dirac
theory. Thus, our task is to show that $\delta x_C^i = \delta x_Q^i$.
To this end we define a set of solutions to the homogeneous equations
obtained by letting ${\cal F}_{LD}^i = 0$ in Eqs.~(\ref{LDone}) and
(\ref{LDtwo}). Thus,
we define the solutions
$(\delta x^i(t),\delta P^i(t)) = (\Delta x_{(j)}^i(t;s),\Delta
P_{(j)}^i(t;s))$, with labels $j=1,2,3$ and $s\in (-\infty,+\infty)$,
to these homogeneous equations
by the following initial conditions:
\beqa
\Delta x_{(j)}^i(s;s) & = & 0 \, , \label{initial1}\\
\Delta P_{(j)}^i(s;s) & = & \delta^{ij}\,. \label{initial2}
\eeqa
Then, the solution  $(\Delta x_{(j)}^i(t;s),\Delta P_{(j)}^i(t;s))$
represents the particle trajectory which coincides with
${\bf x}_0(t)$ at $t=s$ and has an excess
momentum solely in the $j$-direction at this value of $t$.
Then it can readily be seen that
the retarded solution to the inhomogeneous equations (\ref{LDone}) and
(\ref{LDtwo}) is given by
\beqa \label{deltax}
\delta x^i &=& \int^t_{-\infty} ds\, {\cal F}_{\rm LD}^j (s)
\Delta x_{(j)}^{i}(t;s)\,,
\\ \label{deltaP}
 \delta P^i &=& \int^t_{-\infty} ds\, {\cal F}_{\rm LD}^j (s)
\Delta P_{(j)}^{i}(t;s)\,,
\eeqa
where the index $j$ is summed over.
Hence, the classical position shift is
\beq \label{delxC}
\delta x_C^i = \int^0_{-\infty} dt\, {\cal F}_{\rm LD}^j (t) \Delta
x_{(j)}^{i} (0;t)\,.
\eeq

To obtain a similar expression for the quantum position shift $\delta
x_Q^i$ we note first that
\beq
\pd{x^j}{p^i}{t} = \Delta x_{(i)}^j(t;0)\,.
\eeq
This is because
$\epsilon \Delta x_{(i)}^j(t;0)$ with $\epsilon \ll 1$
is the change in the position of the charged particle
in the $j$-direction at time $t$
caused by a change in the $i$-th component of the
momentum at $t=0$ by $\epsilon$.
Thus, the quantum position shift $\delta x_Q^i$ in
Eq.~(\ref{quantshifteqn}) can be written
\beq \label{delxQ}
\delta x_Q^i = - \int_{-\infty}^0 dt\,{\cal F}_{\rm LD}^j(t)
\Delta x_{(i)}^j
(t;0)\,,
\eeq
where the index $j$ is summed over.

The classical and quantum position shifts given by Eqs.~(\ref{delxC}) and
(\ref{delxQ}), respectively, can be shown to be equal
by using conservation of the
symplectic product for the homogeneous equations
(\ref{LDone}) and (\ref{LDtwo}) with ${\cal F}_{\rm LD}^i = 0$.  The
symplectic product between two solutions,
$( \Delta {\bf x}^A,  \Delta {\bf P}^A)$ and
$( \Delta {\bf x}^B,  \Delta {\bf P}^B)$ to these homogeneous equations
is defined by
$\Delta {\bf x}^A\cdot \Delta {\bf P}^B -
\Delta {\bf x}^B \cdot\Delta {\bf P}^A$\,.
It is well known and can easily be verified that the time derivative of
this product vanishes, i.e. that it is conserved.  By equating
the symplectic products of the two solutions
$(\Delta x_{(i)}^k(t;s),\Delta P_{(i)}^k(t;s))$ and
$(\Delta x_{(j)}^k(t;u),\Delta P_{(j)}^k(t;u))$ at $t=s$ and $u$
we have
\beq
\Delta x_{(i)}^k(s;s)\Delta P_{(j)}^k(s;u) -
\Delta x_{(j)}^k(s;u)\Delta P_{(i)}^k(s;s) =
\Delta x_{(i)}^k(u;s)\Delta P_{(j)}^k(u;u) -
\Delta x_{(j)}^k(u;u)\Delta P_{(i)}^k(u;s)\,.
\eeq
This equation and the initial conditions (\ref{initial1}) and
(\ref{initial2}) imply
\beq
-\Delta x_{(j)}^{i}(s;u) = \Delta x_{(i)}^{j}(u;s)\,.
\label{symplecticswap}
\eeq
In particular, we have $-\Delta x_{(j)}^i(0;t) = \Delta x_{(i)}^j(t;0)$.
This equality and Eqs.~(\ref{delxC}) and (\ref{delxQ}) imply that
$\delta x_C^i = \delta x_Q^i$.

\section{Conclusion}\label{conclusion}

In this paper we compared the change in position of a
charged scalar particle due to radiation reaction in both classical and
quantum electrodynamics in three space dimensions. We found that,
for a charged particle accelerated for a finite period of time
by an electromagnetic
potential dependent on one of the space time coordinates, the
change in position in classical electrodynamics using the
Lorentz-Dirac force as a perturbation is the same as that given by
the $\hbar\to 0$ limit of the one-photon-emission calculation in
quantum electrodynamics. This extended the results of Ref.~\cite{HM3}
from one to three dimensions.  Since the forward-scattering amplitude
at order $e^2$ does not contribute to the change in the position in the
$\hbar \to 0$ limit if the potential depends only on
time as shown in Appendix A,
one can conclude that the $\hbar \to 0$ limit of quantum field theory
coincides with the Abraham-Lorentz-Dirac theory at least
in this case.  We expect that
this will be true also with a potential dependent on one space
coordinate.

In our calculations it was crucial that a plane wave corresponded to
a set of parallel classical trajectories with the same motion.
This property was guaranteed by the assumption that
the external electromagnetic potential depended only on one
spacetime coordinate.  If the potential depends on more than
one coordinate,
classical particles with the same initial or final momentum can have
many different motions depending on their initial or final positions.
(Consider, for example, scattering by a central potential.)
It would be interesting if one could overcome this difficulty
and extend our
results to cases with more general potentials.
It would also be interesting to extend our results to radiation
reaction in curved spacetime or to the back-reaction to
gravitational radiation~\cite{Sasaki,Wald}.


\

\

\appendix

\section{Forward-Scattering Amplitude}\label{forscatappendix}

In this Appendix we show that the contribution of
the forward-scattering amplitude at order $e^2$ to the position
change of the particle
vanishes in the $\hbar \to 0$
limit if the external electromagnetic potential depends
only on $t$.  Since this contribution has an
explicit factor of $\hbar$ in Eq.~(\ref{forwardadded}),
we can neglect
any term of order higher than $\hbar^{-1}$ in the forward-scattering
amplitude.
What we demonstrate here is that the one-loop contribution
of order $\hbar^{-1}$ or lower is cancelled exactly
by that from the mass counterterm, which is formally of order
$\hbar^{-2}$.

We use a non-covariant Hamiltonian formulation with the Hamiltonian
density (\ref{Hamiltonian}).  The forward-scattering amplitude
${\cal F}({\bf p})$ for the
state with final momentum ${\bf p}$ is given by
\beqa
2ip_0(2\pi\hbar)^3{\cal F}({\bf p})\delta^3({\bf p}-{\bf p}')
& = & - \frac{i}{\hbar}\int d^4 x \langle 0|A({\bf p}')
{\cal H}_I(x)A^\dagger({\bf p})|0\rangle\nonumber \\
&& - \frac{1}{2\hbar^2}\int d^4x' d^4x\langle 0|A({\bf p}')
T[{\cal H}_I(x'){\cal H}_I(x)]A^\dagger({\bf p})|0\rangle\,,
\eeqa
where $T$ denotes time-ordering, to second order in the
standard time-dependent
perturbation theory in the interaction picture.

We first consider the contribution of the mass counterterm
$-(\delta m^2/\hbar^2)\normal\varphi^\dagger\varphi\normal\,\,$
in the interaction Hamiltonian density
to the forward-scattering amplitude.
This contribution, which we denote by
${\cal F}^{\rm mass}({\bf p})$, can readily be found at first order in
perturbation theory as
\beq\label{masscounter}
{\cal F}^{\rm mass}({\bf p})
= \frac{1}{2\hbar p_0} \int dt\,|\phi_{\bf p}(t)|^2\,\delta m^2\,,
\eeq
where $\phi_{\bf p}(t)$ is the time-dependent part of the scalar mode
function defined by Eq.~(\ref{Phiphi}). (The quantity
${\cal F}^{\rm mass}({\bf p})$ is obviously
divergent in the $t$-integration.
In the end all terms with this divergence property will cancel out.)
The mass parameter in the counterterm in the standard covariant
perturbation theory reads
\beq
\delta m^2  =
\frac{e^2}{\hbar}\int\frac{d^4 q}{(2\pi)^4 i}\left\{
\frac{(p+q)^2}{\left[q^2 - m^2+i\epsilon\right]
\left[(p-q)^2 + i\epsilon\right]} -
\frac{4}{\left[(p-q)^2+i\epsilon\right]}\right\}
\,.
\eeq
It is convenient to integrate over the time component $q_0$ of $q^\mu$
for later purposes.  The result can be given as follows:
\beqa
\delta m^2 & = & \frac{e^2}{\hbar}\int \frac{d^3{\bf q}}{(2\pi)^3}
\left\{
 -\frac{({\bf p}+{\bf q})^2}{4Kq_0}
\left[ \frac{1}{q_0+K-p_0} + \frac{1}{K+p_0+q_0}
\right]\right. \nonumber \\
&& \left. \ \ \ \ \ \ \ \ \ \ \ \  \ \ \ \ \ + \frac{3}{2K} +
\frac{1}{4Kq_0}\left[ \frac{(p_0-q_0)^2}{q_0+K+p_0} +
\frac{(p_0+q_0)^2}{q_0+K-p_0} \right]\right\}
\label{masscounter2}
\eeqa
with  ${\bf K} \equiv {\bf p}-{\bf q}$ and $K \equiv \|{\bf K}\|$,
where $q_0$ is now defined to
be $\sqrt{{\bf q}^2 + m^2}$.

The second term in Eq.~(\ref{Hamiltonian})
contributes to the forward-scattering amplitude at first order in
time-dependent
perturbation theory.  By denoting this contribution by
${\cal F}_2({\bf p})$ we have
\beq
2ip_0(2\pi\hbar)^3{\cal F}_2({\bf p})\delta^3({\bf p}-{\bf p}')
 =  -\frac{ie^2}{\hbar^3}
\sum_{i=1}^3
\int d^4 x\,\langle 0|A({\bf p})A_i(x)A_i(x)
\normal\varphi^\dagger(x)\varphi(x)\normal A^\dagger({\bf
p}')|0\rangle\,.
\eeq
One can readily evaluate this expression with the following result:
\beq
{\cal F}_2({\bf p}) = -\frac{3e^2}{2\hbar^2 p_0}
\int_{-\infty}^{+\infty}
dt\,
|\phi_{\bf p}(t)|^2\int\frac{d^3{\bf q}}{2K(2\pi)^3}\,,
\eeq
where we have used $d^3{\bf k}/k = \hbar^{-2}d^3{\bf q}/K$.

The first term in Eq.~(\ref{Hamiltonian}) contributes to the
forward-scattering amplitude in second order in time-dependent
perturbation theory.  Writing this contribution
as ${\cal F}_1({\bf p})$, we find
\beqa
{\cal F}_1({\bf p})  &  = & \frac{ie^2}{2p_0}
\int \frac{d^3{\bf q}}{2q_0(2\pi\hbar)^3}\frac{1}{2K}\int dt_1
dt_2 \nol
&& \left\{ \theta(t_1-t_2)
\left[\phi_{\bf p}^*(t_1)\phi_{\bf p}(t_2)
\stackrel{\leftrightarrow}{\cal D}_1\hspace{-1mm}
(t_1,t_2,{\bf p},{\bf q})
\phi_{\bf q}(t_1)\phi_{\bf q}^*(t_2)\right]e^{-iK(t_1-t_2)/\hbar}
\right. \nol
&& + \left. \theta(t_2-t_1)
\left[\phi_{\bf p}^*(t_1)\phi_{\bf p}(t_2)
\stackrel{\leftrightarrow}{\cal D}_1\hspace{-1mm}
(t_1,t_2,{\bf p},{\bf q})
\phi_{\bf q}^*(t_1)\phi_{\bf q}(t_2)\right]
e^{iK(t_1-t_2)/\hbar}\right\}\,,\label{firstF1}
\eeqa
where
\beq
\stackrel{\leftrightarrow}{\cal D}_1\hspace{-1mm}
(t_1,t_2,{\bf p},{\bf q}) \equiv
 - \hbar^2\stackrel{\leftrightarrow}{\partial}_{t_1}
\stackrel{\leftrightarrow}{\partial}_{t_2}
+ \left[{\bf p}+{\bf q}-2{\bf V}(t_1)\right]\cdot
\left[{\bf p}+{\bf q}-2{\bf V}(t_2)\right]
\,. \label{defD1}
\eeq
It is convenient to define the variables $t$ and $\eta$
by $t_1 = t-\hbar\eta/2$ and $t_2 = t+ \hbar\eta/2$.
Then
\beq  \label{momint}
{\cal F}_1({\bf p}) =  \frac{ie^2}{2\hbar^2p_0}
\int \frac{d^3{\bf q}}{2q_0(2\pi)^3}\frac{1}{2K}
\int dt \left[ G_{-}({\bf p},{\bf q},t) + G_{+}
({\bf p},{\bf q},t)\right]\,,
\eeq
with
\beqa
G_{-}({\bf p},{\bf q},t) & = & \int_{-\infty}^0 d\eta
\left[ \phi_{\bf p}^*(t_1)\phi_{\bf p}(t_2)
\stackrel{\leftrightarrow}{\cal D}_2\hspace{-1mm}
(t,\eta,{\bf p},{\bf q})
\phi_{\bf q}(t_1)\phi_{\bf q}^*(t_2)\right] e^{iK\eta}\,, \\
G_{+}({\bf p},{\bf q},t) & = & \int_0^{\infty}d\eta
\left[ \phi_{\bf p}^*(t_1)\phi_{\bf p}(t_2)
\stackrel{\leftrightarrow}{\cal D}_2\hspace{-1mm}
(t,\eta,{\bf p},{\bf q})
\phi_{\bf
q}^*(t_1)\phi_{\bf q}(t_2)\right] e^{-iK\eta}\,,
\eeqa
where
\beq
\stackrel{\leftrightarrow}{\cal D}_2\hspace{-1mm}
(t,\eta,{\bf p},{\bf q})
\equiv - \frac{\hbar^2}{4}\stackrel{\leftrightarrow}{\partial}_{t}^2
-\stackrel{\leftrightarrow}{\partial}_{\eta}^2
+ \left[{\bf p}+{\bf q}-2{\bf V}(t_1)\right]\cdot
\left[{\bf p}+{\bf q}-2{\bf V}(t_2)\right]
\,.
\eeq

We write the time-dependent part of the mode
function, $\phi_{\bf q}(t)$, as
\beq
\phi_{\bf q}(t) = \sqrt{\frac{q_0}{\sigma_{\bf q}(t)}}\psi_{\bf q}(t)
\exp \left[ - \frac{i}{\hbar}\int_0^t \sigma_{\bf q}(t')\,dt'\right]\,,
\eeq
where $\psi_{\bf q}(t) \to 1$ as $\hbar \to 0$.
By substituting this expression in Eq.~(\ref{phiequation})
we can expand $\psi_{\bf q}(t)$ in terms of $\hbar$ as
\beq
\psi_{\bf q}(t) = 1 + i\hbar \psi^{(1)}_{\bf q}(t) + {\cal
O}(\hbar^2)\,.
\eeq
The explicit form of $\psi^{(1)}_{\bf q}(t)$ is unnecessary though it
can easily be found.  Note also that
\beq
\phi_{\bf q}^*(t_1)\phi_{\bf q}(t_2)
= \frac{\psi_{\bf q}^*(t_1)\psi_{\bf q}(t_2)}
{\sqrt{\sigma_{\bf q}(t_1)\sigma_{\bf q}(t_2)}}
\exp\left[ -i\int_{-\eta/2}^{\eta/2}
\sigma_{\bf q}(t+\hbar\zeta)\,d\zeta\right]\,.
\eeq
It can readily be shown
that the functions $G_{\pm}({\bf p},{\bf q},t)$ are of the form
\beqa
G_{\pm} ({\bf p},{\bf q},t)
& = & \pm\int_{0}^{\pm\infty}d\eta \left[ f_{\pm}({\bf
p},{\bf q},t) + {\cal O}(\hbar^2)\right]\nol
&& \ \ \ \ \ \ \ \ \ \ \times \exp\left\{ \mp
i\int_{-\eta/2}^{+\eta/2} d\zeta \left[\pm \sigma_{\bf
p}(t+\hbar\zeta)
+ \sigma_{\bf q}(t+\hbar\zeta) + K\right]\right\}\,. \label{Gpm}
\eeqa
where the function $f_{\pm}({\bf p},{\bf q},t)$ can be found as
\beq\label{fplus}
f_{\pm}({\bf p},{\bf q},t)  = \left\{- \[
\sigma_{\bf p}(t)\mp\sigma_{\bf q}(t)\]^2
+ \[{\bf p}+{\bf q}-2{\bf V}(t)\]^2\right\}|\phi_{\bf p}(t)|^2
|\phi_{\bf q}(t)|^2\,.
\eeq
An important fact to note here
is that there are no terms of order $\hbar$ in the
pre-factor inside the first square brackets in Eq.~(\ref{Gpm}).  This fact is a simple
consequence of the equations
$\psi_{\bf q}^*(t_1)\psi_{\bf q}(t_2) = 1 + {\cal O}(\hbar^2)$ and
\beq
\left[{\bf p}+{\bf q}-{\bf V}(t_1)\right]\cdot
\left[{\bf p}+{\bf q}-{\bf V}(t_2)\right]
= \left[{\bf p}+ {\bf q}-{\bf V}(t)\right]^2 + {\cal O}(\hbar^2)\,.
\eeq

Let us first consider the integral $G_{+}({\bf p},{\bf q},t)$.
We change the integration
variable from $\eta$ to $\beta$ defined by the following
relation:
\beq\label{defbeta}
[\sigma_{\bf p}(t) + \sigma_{\bf q}(t) + K]\beta\equiv
\int_{-\eta/2}^{\eta/2}
\left[\sigma_{\bf p}(t+\hbar\zeta)
+ \sigma_{\bf q}(t+\hbar\zeta) + K\right]\,d\zeta\,.
\eeq
Expanding the integrand and integrating, we solve
for $\eta$ as a function of $\beta$ for small $\hbar$ and find
\beq
\eta = \left[ 1 - \frac{1}{24}\frac{\ddot{\sigma}_{\bf p}(t)+
\ddot{\sigma}_{\bf q}(t)}
{\sigma_{\bf p}(t)+\sigma_{\bf q}(t)+K}\hbar^2\beta^2
+ {\cal O}(\hbar^4\beta^4)\right]\beta  \label{tau}
\eeq
and
\beq
d\eta = \left[ 1 - \frac{1}{8}\frac{\ddot{\sigma}_{\bf p}(t)
+\ddot{\sigma}_{\bf q}(t)}
{\sigma_{\bf p}(t)+\sigma_{\bf q}(t)+K}\hbar^2\beta^2
+ {\cal O}
(\hbar^4\beta^4)\right] d\beta\,.
\eeq
Then we have
\beq\label{Gplus}
G_{+}({\bf p},{\bf q},t) = \int_{0}^{\infty}d\beta
\left[ f_{+}({\bf p},{\bf q},t) + {\cal O}(\hbar^2)\right]\,
\exp\left\{ -i\left[\sigma_{\bf p}(t)
+ \sigma_{\bf q}(t) + K\right]\beta\right\}\,.
\eeq
Introducing a convergence factor by replacing
$K$ by $K-i\epsilon$, we obtain
\beq
G_{+}({\bf p},{\bf q},t) = -\frac{if_{+}({\bf p},{\bf q},t)}
{\sigma_{\bf q}(t)+\sigma_{\bf p}(t)+K} + {\cal O}(\hbar^2)\,.
\eeq
The corresponding term in the forward-scattering amplitude is
\beq
{\cal F}_{1+}({\bf p})
=  \frac{e^2}{2\hbar^2 p_0}\int dt
\int \frac{d^3{\bf q}}{2q_0(2\pi)^3}\frac{1}{2K}
\frac{f_{+}({\bf p},{\bf q},t)}{\sigma_{\bf p}(t)
+ \sigma_{\bf q}(t) + K} + {\cal O}(\hbar^0)\,,  \label{div1}
\eeq
The amplitude ${\cal F}_{1+}({\bf p})$
can readily be seen to be ultra-violet divergent.

Next we analyze the contribution from $G_{-}({\bf p},{\bf q},t)$.
One cannot proceed as above because of the infrared divergence
in the ${\bf q}$-integration as we shall see.
We define the variable $\tilde{\beta}$ in analogy with the variable
$\beta$ in Eq.~(\ref{defbeta}) by
\beq
[-\sigma_{\bf p}(t) + \sigma_{\bf q}(t) + K]\tilde{\beta} \equiv
\int_{-\eta/2}^{\eta/2}
\left[-\sigma_{\bf p}(t+\hbar\zeta)
+ \sigma_{\bf q}(t+\hbar\zeta) + K\right]\,d\zeta\,.
\eeq
Now, for small
$K = \|{\bf p} - {\bf q}\|$, we have
\beq\label{smallK}
-\sigma_{\bf p}(t)+\sigma_{\bf q}(t) + K \approx
 K - {\bf v}(t)\cdot {\bf K}\,,
\eeq
where ${\bf v}(t) = [{\bf p}-{\bf V}(t)]/\sigma_{\bf p}(t)$ is
the velocity of the classical particle with final momentum ${\bf p}$
(see Eq.~(\ref{wkbvel})).
Hence, in the limit $K\to 0$ one finds
\beq
\tilde{\beta} = \frac{1}{1-{\bf v}(t)\cdot{\bf n}}
\int_{-\eta/2}^{\eta/2} \[1-{\bf v}(t+\hbar\zeta)\cdot{\bf n}
\]d\zeta\,,
\eeq
where ${\bf n} \equiv {\bf K}/K$.  Thus, if we write
$d\eta = J({\bf p},{\bf q},t,\hbar\tilde{\beta})d\tilde{\beta}$, then
the function $J({\bf p},{\bf q},t,\hbar\tilde{\beta})$ is finite
as $K\to 0$.
Hence, the expression
corresponding to Eq.~(\ref{Gplus}) can be given in the following form:
\beqa
G_{-}({\bf p},{\bf q},t)
& = & \int_{-\infty}^{0} d\tilde{\beta}
\left[ f_{-}({\bf p},{\bf q},t)
+ \sum_{n,d}\hbar^n\tilde{\beta}^d f_{nd-}({\bf p},{\bf q},t)\right]
\exp\left\{ i\left[ -\sigma_{\bf p}(t) + \sigma_{\bf q}(t) + K\right]
\tilde{\beta}\right\}\nol
& = & \frac{-if_{-}({\bf p},{\bf q},t)}
{-\sigma_{\bf p}(t)+ \sigma_{\bf q}(t)+K}
+ \sum_{n,d}\frac{(-i)^d d!\hbar^n f_{nd-}({\bf p},{\bf q},t)}
{\left[-\sigma_{\bf p}(t)+\sigma_{\bf q}(t)+K\right]^{d+1}}
 \label{hexp}
\eeqa
with $n\geq 2$ and $n\geq d\geq 0$,
where $f_{nd-}({\bf p},{\bf q},t)$ are finite as
$K \to 0$.

Now, if we substitute Eq.~(\ref{hexp})
in Eq.~(\ref{momint}), then the terms
with $d \geq 1$ are infrared divergent in the ${\bf q}$-integration
because $\lim_{K\to 0}[-\sigma_{\bf p}(t)+\sigma_{\bf q}(t)+K] \to 0$
as can be seen from Eq.~(\ref{smallK}).
(Note that $d^3{\bf q} = d^3{\bf K}$.)
To deal with this
difficulty we cut off the integral over ${\bf q}$ by
requiring $K \geq K_{0} = \hbar^\alpha\lambda$ with
$\frac{3}{4} < \alpha < 1$,
where $\lambda$ is a positive constant, and postpone the analysis of
the contribution
from $K \leq K_{0}$ till later. (The reasoning for this particular
choice of limits for $\alpha$ will be seen later). Then, we find that the
small-$K$ contribution of each term in Eq.~(\ref{hexp}) to the
${\bf q}$-integral
behaves like $\hbar^{n}K_0^{1-d} = \hbar^{n + (1-d)\alpha}\lambda^{1-d}$
if $d \geq 2$
and $\hbar^n \log(\hbar^\alpha \lambda)$ if $d=1$.
Now, since
$1-\alpha > 0$, $n \geq 2$ and $n\geq d$, we have $n + (1-d)\alpha \geq
2-\alpha$. Therefore, in the $\hbar\to 0$ limit, the $f_{nd-}$ terms
will not contribute above the cut-off, leaving only the first term.
Let us thus combine the leading order terms from ${\cal F}_{1}$
(using $G_{+}$ and the first term of $G_{-}$, above and below the cut-off) and the result for ${\cal
F}_2$ to define
 \beqa
{\cal F}^{\rm leading}({\bf p})
& = &  \frac{\hbar^{-2}e^2}{2p_0}
\int dt\,|\phi_{\bf p}(t)|^2
\int \frac{d^3{\bf q}}{2\sigma_{\bf q}(t)(2\pi)^3}\frac{1}{2K}
\nol
&& \times \left\{ - 6\sigma_{\bf q}(t) +
\frac{-\left[\sigma_{\bf p}(t)+\sigma_{\bf q}(t)\right]^2
+ \[{\bf p}+{\bf q}-2{\bf V}(t)\]^2}
{\sigma_{\bf q}(t) + K - \sigma_{\bf p}(t)}  \right.\nol
&& \left. \ \ \ \ \ \ \
+ \frac{
- \left[\sigma_{\bf p}(t)-\sigma_{\bf q}(t)\right]^2
+ \[{\bf p}+{\bf q}-2{\bf V}(t)\]^2}{\sigma_{\bf q}(t)+K +
\sigma_{\bf p}(t)}\right\}\,,\label{leading}
\eeqa
(For $f_{\pm}$ we have used the relation $|\phi_{\bf q}|^2=q_0/\sigma_{\bf
q}(t) \[ 1+ {\cal O}(\hbar^2) \]$).
What remains is the contribution of $G_{-}$, minus the first term (which is in ${\cal F}^{\rm
leading}$),
from below the cut-off. Thus we define
\beqa
{\cal F}^{<}({\bf p}) & = & \frac{ie^2}{2p_0}
\int_{K \leq \hbar^\alpha\lambda}
\frac{d^3{\bf q}}{2q_0(2\pi\hbar)^3}\frac{1}{2K}\int dt_1 dt_2
\theta(t_1-t_2)e^{-iK(t_1-t_2)/\hbar}\nol
&& \ \ \ \ \ \times \phi_{\bf p}^*(t_1)\phi_{\bf p}(t_2)
\stackrel{\leftrightarrow}{\cal D}_1\hspace{-1mm}
(t_1,t_2,{\bf p},{\bf q})
\phi_{\bf q}(t_1)\phi_{\bf q}^*(t_2)
\\
{\cal F}^{<,0}({\bf p}) & = &
 \frac{\hbar^{-2}e^2}{2p_0}
\int dt\,|\phi_{\bf p}(t)|^2 \int_{K \leq \hbar^\alpha\lambda}
 \frac{d^3{\bf q}}{2\sigma_{\bf q}(t)(2\pi)^3}
\frac{1}{2K} \nol
&& \ \ \ \ \ \times
\frac{ - \left[\sigma_{\bf p}(t)+\sigma_{\bf q}(t)\right]^2
+ \[{\bf p}+{\bf q}-2{\bf V}(t)\]^2}
{\sigma_{\bf q}(t)+K -\sigma_{\bf p}(t)}\,,\label{restrest}
\eeqa
with the operator
$\stackrel{\leftrightarrow}{\cal D}_1\hspace{-1mm}
(t_1,t_2,{\bf p},{\bf q})$
defined by Eq.~(\ref{defD1}). Hence the forward-scattering amplitude
can now be written as follows:
\beq
{\cal F}({\bf p})  =  {\cal F}^{\rm mass}({\bf p})
+ {\cal F}^{\rm leading}({\bf p})
+ {\cal F}^{<}({\bf p})  - {\cal F}^{<,0}({\bf p})
+ {\cal O}(\hbar^{-\alpha})\,,
\label{fourterms}
\eeq
with the contribution from the mass counterterm,
${\cal F}^{\rm mass}({\bf p})$, given by Eq.~(\ref{masscounter}).
Below we show the equality
${\cal F}^{\rm leading}({\bf p}) = -{\cal F}^{\rm mass}({\bf p})$ and
demonstrate
that the quantity
${\cal F}^{<}({\bf p}) - {\cal F}^{<,0}({\bf p})$ is of
order $\hbar^{-1}$ and is purely
imaginary at this order.  Then one can conclude that
the real part of the forward-scattering amplitude is of order higher
than $\hbar^{-1}$, and hence does
not contribute to the position change, as claimed in this paper.

The equality
${\cal F}^{\rm leading}({\bf p}) = - {\cal F}^{\rm mass}({\bf p})$
can readily be obtained by letting
$\tilde{{\bf q}} \equiv {\bf q} - {\bf V}(t)$ and
$\tilde{{\bf p}} \equiv {\bf p} - {\bf V}(t)$ in
Eq.~(\ref{leading}).
Then writing
$\sigma_{\bf p}(t) = \sqrt{\tilde{\bf p}^2 + m^2} = \tilde{p}_0$
and
$\sigma_{\bf q}(t) = \sqrt{\tilde{\bf q}^2 + m^2} = \tilde{q}_0$,
we have
\begin{align}
{\cal F}^{{\rm leading}}({\bf p})
= & \frac{\hbar^{-2}e^2}{2p_0}
\int dt\,|\phi_{\bf p}(t)|^2 \int
\frac{d^3\tilde{{\bf q}}}{2\tilde{q}_0(2\pi)^3}\frac{1}{2K}\nol
&\times \left\{ -6\tilde{q}_0 +
\frac{-\left(\tilde{p}_0 +\tilde{q}_0\right)^2
+ (\tilde{{\bf p}}+\tilde{{\bf q}})^2}
{\tilde{q}_0 + K - \tilde{p}_0}
+ \frac{
- \left(\tilde{p}_0 - \tilde{q}_0\right)^2
+ (\tilde{{\bf p}}+\tilde{{\bf q}})^2}{\tilde{q}_0+K + \tilde{p}_0}
\right\} \,.
\end{align}
This can be seen to be the
negative of ${\cal F}^{\rm mass}({\bf p})$ in
Eq.~(\ref{masscounter}) with $\delta m^2$ given by
Eq.~(\ref{masscounter2}) with ${\bf q}$ and ${\bf p}$ replaced by
$\tilde{\bf q}$ and $\tilde{\bf p}$, respectively.  Clearly, one
can freely rename the integration variable $\tilde{\bf q}$ as
${\bf q}$.  Also, it is well known that $\delta m^2$
does not explicitly depend on
${\bf p}$, but only on $p_0^2 - {\bf p}^2 = m^2$.  Since we have
$\tilde{p}_0^2 - \tilde{\bf p}^2 = m^2$,
we may replace $\tilde{\bf p}$ by ${\bf p}$.
Hence we can conclude that
${\cal F}^{\rm leading}({\bf p}) = -{\cal F}^{\rm mass}({\bf p})$.

Next we analyze the contribution ${\cal F}^{<}({\bf p})$.
By changing the integration variable from ${\bf q}$ to
${\bf K} = {\bf p}-{\bf q}$ and then to ${\bf k} = {\bf K}/\hbar$,
we find
\beqa
 {\cal F}^{< }({\bf p})
& = & \frac{ie^2}{2\hbar p_0}
\int dt_1 dt_2\int_{k \leq \hbar^{\alpha-1}\lambda}
\frac{d^3{\bf k}}{2q_0(2\pi)^3}\frac{1}{2k}\theta(t_1-t_2)\nol
&&\times
\left[\phi_{\bf p}^*(t_1)\phi_{\bf p}(t_2)
\stackrel{\leftrightarrow}{\cal D}_1\hspace{-1mm}
(t_1,t_2,{\bf p},{\bf q})
\phi_{\bf q}(t_1)\phi_{\bf q}^*(t_2)\right]e^{-ik(t_1-t_2)}\,.
\eeqa
Now we have ${\bf q}= {\bf p} - \hbar {\bf k}$.  (Note also that
the upper limit $\hbar^{\alpha-1}\lambda$ of integration for $k$
becomes infinite as $\hbar$ tends to zero.)  Hence,
we have ${\bf q} \to {\bf p}$ for all ${\bf k}$ as $\hbar\to 0$ because
$\hbar\cdot\hbar^{\alpha-1}\lambda \to 0$.
The exponential factor takes the form
\begin{multline}
\exp\left\{ i\int_0^t d\zeta\left[K + \sigma_{\bf
q}(\zeta)-\sigma_{\bf p} (\zeta)\right]/\hbar\right\} \\ =
\exp\left\{ i\int_0^t \left[ k - \frac{\partial{\sigma}_{\bf
p}(\zeta)}{\partial {\bf p}}\cdot{\bf k} +
\frac{1}{2}\frac{\partial^2\sigma_{\bf p}(\zeta)}{\partial
p^i\partial p^j}\hbar k^ik^j + \cdots\right]d\zeta\right\}\,.
\end{multline}
Thus, to truncate the series in the exponent at the second term
for all ${\bf k}$ in the integration range, we need
$\hbar (\hbar^{\alpha-1})^2 \to 0$ as $\hbar\to 0$.  This is satisfied
due to the requirement $\alpha >\frac{3}{4}$.  Thus we have
\begin{align}
{\cal F}^{< }({\bf p})
 = & \frac{ie^2}{\hbar}
\int dt_1 dt_2\int_{k \leq \hbar^{\alpha-1}\lambda} \frac{d^3{\bf
k}}{2k(2\pi)^3} \theta(t_1-t_2)\left\{ - 1  +
\frac{\left[{\bf p}-{\bf V}(t_1)\right]\cdot \left[ {\bf p}-{\bf
V}(t_2)\right]}{\sigma_{\bf p}(t_1)\sigma_{\bf p}(t_2)} \right\}\nol
& \times\exp
\left[ i k(t_2-t_1) - \int_{t_1}^{t_2}\frac{{\bf p}-{\bf
V}(\zeta)}{\sigma_{\bf p}(\zeta)}\cdot {\bf k} d\zeta\right]
+ {\cal O}(\hbar^{4\alpha-4})\,.
\end{align}
Since $4\alpha -4 > -1$ because of
the requirement $\alpha >\frac{3}{4}$, the
non-leading terms do not contribute to the position shift in the limit
$\hbar \to 0$.  We drop this contribution from now on for this
reason.
Recalling that $[{\bf p}-{\bf V}(t)]/\sigma_{\bf p}(t)$ is the velocity
of the corresponding classical particle, $d{\bf x}/dt$, we obtain
at leading order, in analogy with Eq.~(\ref{Aiai}),
\beq
{\cal F}^{<}({\bf p}) = \frac{ie^2}{\hbar}\int_{k \leq
\hbar^{\alpha-1}\lambda}
\frac{d^3{\bf k}}{(2\pi)^32k}\int_{-\infty}^{+\infty}d\xi
\int_{-\infty}^{+\infty}d\xi'
\theta(\xi'-\xi)\frac{dx^\mu}{d\xi}\frac{dx_\mu}{d\xi'}
e^{ik(\xi'-\xi)}\,,
\eeq
where we have defined $\xi \equiv t_1 - {\bf n}\cdot {\bf x}(t_1)$ and
$\xi' \equiv t_2 - {\bf n}\cdot{\bf x}(t_2)$ with
${\bf n} \equiv {\bf k}/k$. If we write the Heaviside function as
$\theta(\xi-\xi')=1/2 + \epsilon(\xi-\xi')/2$, where $\epsilon(\xi'-\xi)=1$ if $\xi'>\xi$ and
$\epsilon(\xi'-\xi)=-1$ if $\xi'< \xi$,
then the first and second terms give
the imaginary and real parts of ${\cal F}^{<}({\bf p})$, respectively.
Thus, twice the imaginary part is
\beq
2\,{\rm Im}\,{\cal F}^{<}({\bf p}) = \frac{ie^2}{\hbar}\int_{k \leq
\hbar^{\alpha-1}\lambda}
\frac{d^3{\bf k}}{(2\pi)^32k}\int_{-\infty}^{+\infty}d\xi
\int_{-\infty}^{+\infty}d\xi'
\frac{dx^\mu}{d\xi}\frac{dx_\mu}{d\xi'}
e^{ik(\xi'-\xi)}\,.\label{imaginary}
\eeq
This coincides with the emission probability found in Ref.~\cite{HM3}
after we let $\hbar^{\alpha-1}\lambda \to \infty$
as required by unitarity.  The real part is given by
\beq
{\rm Re}\, {\cal F}^{<}({\bf p})
 =  \frac{ie^2}{2\hbar}\int_{k \leq \hbar^{\alpha-1}\lambda}
\frac{kdkd\Omega}{2(2\pi)^3}
\int_{-\infty}^{+\infty}
d\xi \int_{-\infty}^{+\infty} d\xi'
\epsilon(\xi'-\xi)
\frac{dx^\mu}{d\xi}\chi(\xi)\frac{dx_\mu}{d\xi'}\chi(\xi')
e^{ik(\xi'-\xi)}\,,
\eeq
We have introduced the cut-off function $\chi(\xi)$ as in
Eq.~(\ref{cut-off}). Now, we integrate by parts with respect to
the variable $\xi$ and add the result of integrating by parts with
respect to $\xi'$, then divide by two. Thus we find
\begin{multline}
{\rm Re}\,{\cal F}^{<}({\bf p})  =
 - \frac{e^2}{4\hbar}\int_{k\leq \hbar^{\alpha-1}\lambda}
\frac{dkd\Omega}{2(2\pi)^3}
\int_{-\infty}^{+\infty}
d\xi \int_{-\infty}^{+\infty} d\xi'
\left\{ 4\delta(\xi'-\xi) \frac{dx^\mu}{d\xi'}\frac{dx_\mu}{d\xi}\right.
\\ \left.
+ \epsilon(\xi'-\xi)\left[ \(\frac{d}{d\xi'}-\frac{d}{d\xi}\) \frac{dx^\mu}{d\xi'}
\chi(\xi')\frac{dx_\mu}{d\xi}\chi(\xi)\right]
\right\}e^{ik(\xi'-\xi)}\,.
\end{multline}
The factor multiplying
$e^{ik(\xi'-\xi)}$ in the second term is symmetric in $\xi$ and $\xi'$.
Hence the $k$-integral of $e^{ik(\xi'-\xi)}$ can be made into
$\delta(\xi'-\xi)$ by extending the integration range of $k$ to
$(-\infty,+\infty)$ and dividing by two in the limit
$\hbar \to 0 $. The factor inside the square brackets becomes
zero if one lets $\xi'=\xi$.  Hence the contribution to
${\rm Re}\,{\cal F}^{<}({\bf p})$ from the second term
is of order higher than $\hbar^{-1}$.
Evaluating the first term with the use of
$(dx^\mu/dt)(dx_\mu/dt)=1-{\bf v}^2$ and
$\dot{\xi} = 1-{\bf n}\cdot{\bf v}$, we find
\beq
{\rm Re}\,{\cal F}^{<}({\bf p}) =
- \frac{e^2\lambda}{16\pi^3\hbar^{2-\alpha}}\int_{-\infty}^{+\infty}
dt \int d\Omega \frac{1-{\bf v}^2}{1-{\bf n}\cdot {\bf v}}
\label{OK}
\eeq
to order $\hbar^{-1}$.
Now, let us consider ${\cal F}^{<,0}({\bf p})$. By substituting the
small-$K$ approximation (\ref{smallK}) in Eq.~(\ref{restrest}) and
noting the equations $|\phi_{\bf p}(t)|^2 = p_0/\sigma_{\bf p}(t)$,
$\sigma_{\bf p}(t) = m/\sqrt{1-{\bf v}^2}$,
$[{\bf p}-{\bf V}(t)]/\sigma_{\bf p}(t) = {\bf v}$ and
$[\sigma_{\bf p}(t)]^2 - [{\bf p}-{\bf V}(t)]^2 = m^2$, we
find that ${\cal F}^{<,0}({\bf p})$ is indeed equal to
the leading term of ${\rm Re}\,{\cal F}^{<}({\bf p})$
given by Eq.~(\ref{OK}).

Thus, we have shown the equality
${\cal F}^{\rm mass}({\bf p})+{\cal F}^{\rm leading}({\bf p}) = 0$ and
demonstrated
that ${\cal F}^{<}({\bf p}) - {\cal F}^{<,0}({\bf p})$ is
of order $\hbar^{-1}$ but is purely imaginary at this order.
Therefore, from Eq.~(\ref{fourterms}) we conclude that
the real part of the forward-scattering amplitude,
${\rm Re}\,{\cal F}({\bf p})$, is of order higher than $\hbar^{-1}$,
and hence does not contribute to the change of position of the
scalar particle given by Eq.~(\ref{forwardadded}).


\begin{thebibliography}{99}

\bibitem{Abraham} M.\ Abraham and R.\ Becker,
 {\it Theorie der Elektrizit\"at},
Vol.\ II, (Springer, Leipzig, 1933).

\bibitem{Lorentz} H.\ A.\ Lorentz, {\it Theory of electrons}, (Dover,
New York, 1952).

\bibitem{Dirac} P.\ A.\ M.\ Dirac,
Proc.\ Roy.\ Soc. London {\bf A167}, 148 (1938).



\bibitem{Teit} C.\ Teitelboim,
Phys.\ Rev.\ D\ {\bf 1}, 1572 (1970); {\it ibid}. {\bf 3}, 297
(1971); {\it ibid}. {\bf 4}, 345 (1971).

\bibitem{Poisson} E.\ Poisson, {\it An introduction to the
Lorentz-Dirac equation}, arXiv:gr-qc/9912045.

\bibitem{Rohrlich} F.\ T.\ Rohrlich,
{\it Classical charged particles}, (Addison-Wesley, Reading,
Mass., 1965).

\bibitem{Jackson} J.\ D.\ Jackson,
{\it Classical electrodynamics}, (Wiley, New York, 1975).

\bibitem{Landau} L.\ D.\ Landau and E.\ M.\ Lifshitz, {\it The
classical theory of fields}, (Pergamon, Oxford, 1962).

\bibitem{WE} E.\ Flanagan and R.\ M.\ Wald, Phys.\ Rev.\ D\ {\bf 54},
6233 (1996), arXiv: gr-qc/9602052.

\bibitem{Higuchi2} A.\ Higuchi,
arXiv: quant-ph/9812036; Phys.\ Rev.\ D {\bf 66}, 105004 (2002);
Erratum {\em ibid.} {\bf 69}, 129903 (2004), arXiv:
quant-ph/0208017.

\bibitem{HM2} A.\ Higuchi and G.\ D.\ R.\ Martin, Phys.\ Rev.\ D,
{\bf 70}, 081701(R) (2004), arXiv: quant-ph/0407162.

\bibitem{HM3} A.\ Higuchi and G.\ D.\ R.\ Martin, (2004) arXiv:
quant-ph/0501026, to appear in Foundations of Physics.

\bibitem{MS} E.\ J.\ Moniz and D.\ H.\ Sharp,
Phys.\ Rev.\ D {\bf 10}, 1133 (1974); {\it ibid}. {\bf 15}, 2850
(1977).

\bibitem{beilok} P.\ R.\ Johnson and B.\ L.\ Hu,
Phys.\ Rev.\ D\ {\bf 65}, 065015 (2002), arXiv: quant-ph/0101001.

\bibitem{Tsyt} V.\ S.\ Krivitski\v{\i} and V.\ N. Tsytovich,
Sov.\ Phys.\ Usp.\ {\bf 34}, 250 (1991).

\bibitem{Newton-Wigner} T.\ D.\ Newton and E.\ P.\ Wigner,
Rev.\ Mod.\ Phys.\ {\bf 21}, 400 (1949).

\bibitem{Sasaki} Y.\ Mino, M.\ Sasaki and T.\ Tanaka,
Phys.\ Rev.\ D {\bf 55}, 3457 (1997), arXiv: gr-qc/9606018.

\bibitem{Wald} T.\ C.\ Quinn and R.\ M.\ Wald,
Phys.\ Rev.\ D {\bf 56}, 3381 (1997), arXiv: gr-qc/9610053.

\end{thebibliography}
\end{document}